\documentclass[11pt,epsfig]{article}

\usepackage{amssymb}
\usepackage{color}
\usepackage{amsmath}
\usepackage{graphicx}
\begin{document}
\newcommand{\be}{\begin{eqnarray}}

\newcommand{\ee}{\end{eqnarray}}
\title{{\bf Quantization of the interior Schwarzschild black hole }}
\author{Shahram Jalalzadeh$^{1}$\thanks{s-jalalzadeh@sbu.ac.ir}\,\, and\,\,\,
Babak
Vakili$^{2}$\thanks{b-vakili@iauc.ac.ir}\,\,\,\\\\$^1${\small {\it
Department of Physics, Shahid Beheshti University, G. C. Evin,
Tehran 19839, Iran}}\\$^2${\small {\it Department of Physics,
Islamic Azad University, Chalous Branch, P.O. Box 46615-397,
Chalous, Iran}} $^{}${}}
\date{}
\maketitle

\begin{abstract}
We study a Hamiltonian quantum formalism of a spherically
symmetric space-time which can be identified with the interior of
a Schwarzschild black hole. The phase space of this model is
spanned by two dynamical variables and their conjugate momenta. It
is shown that the classical Lagrangian of the model gives rise the
interior metric of a Schwarzschild black hole. We also show that
the the mass of such a system is a Dirac observable and then by
quantization of
the model by Wheeler-DeWitt approach and constructing suitable wave packets we get the mass spectrum of the black hole. \vspace{5mm}\noindent \vspace{0.8mm}\newline
\end{abstract}

\section{Introduction}
Black hole physics has played a central role in conceptual
discussion of general relativity in classical and quantum levels.
For example regarding event horizons, space-time singularities and
also studying the aspects of quantum field theory in curved
space-time. In classical point of view, the horizon of a black
hole which is a one way membrane, and the space-time singularities
are some interesting features of the black hole solutions in
general relativity \cite{1}. In spirit of the Ehrenfest principle,
any classical adiabatic invariant corresponds to a quantum entity
with discrete spectrum, Bekenstein conjectured that the horizon
area of a non extremal quantum black hole should have a discrete
eigenvalue spectrum \cite{Bek}. Also, the black hole
thermodynamics is based on applying quantum field theory to the
curved space-time of a black hole \cite{Haw}. According to this
formalism, the Hawking radiation of a black hole is due to random
processes in the quantum fields near the horizon. The mechanism of
this thermal radiation can be explained in terms of pair creation
in the gravitational potential well of the black hole \cite{Brl}.
The conclusions of the above works are that the temperature of a
black hole is proportional to the surface gravity and that the
area of its event horizon plays the role of its entropy. In this
scenario, the black hole is akin to a thermodynamical system
obeying the usual thermodynamic laws, often called the laws of
black hole mechanics, first formulated by Hawking \cite{Haw}. In
more recent times, this issue has been at the center of concerted
efforts to describe and make clear various aspects of the problem
that still remain unclear, for a review see \cite{Pad}. With the
birth of string theory \cite{Seib}, as a candidate for quantum
gravity and loop quantum gravity \cite{Asht}, a new window was
opened to the problem of black hole radiation. This was because
the nature of black hole radiation is such that quantum gravity
effects cannot be neglected \cite{Muk}. According to all of the
above remarkable works, it is believed that a black hole is a
quantum mechanical system and thus like any other quantum system
its physical states can be described by a wavefunction. Indeed,
due to its fundamental conceptual role in quantum general
relativity, we may use it as a starting point for testing
different constructions of quantum gravity \cite{2,Kief}.

On the other hand, studying the interior of the Schwarzschild
black hole is interesting for various reasons \cite{Dor}. Many
authors expressed the idea that the interior of a black hole can
be considered as an anisotropic cosmological model \cite{s1}. In
this direction, the author of \cite{s2} used  it as a cosmological
model to describe time dependent cosmological constant. One of the
most important and interesting cosmological usage of the interior
solution is considering black hole as a spawn of mother universe
so that our universe was born from it as a daughter universe
\cite{s3}. Many cosmological models in which our universe emerges
from the interior of a black hole have been proposed in \cite{s4}.
From a mathematical point of view, the space-time metric of the
interior of a black hole can be constructed from its exterior
metric. Indeed, one of the interesting features of general
relativity is to generate cosmological solutions from known static
solutions of the Einstein field equations. The methods from which
one finds such solutions are investigated in \cite{Que}.
Generally, these methods are based on a diffeomorphism between the
known static solutions and the corresponding cosmological model.
For example, the coordinate transformation $t \leftrightarrow r$
converts the Schwarzschild metric to a dynamical model which can
be identified with its inside space-time. Indeed, at the horizon
of a Schwarzschild black hole the light cone tips over and for
$r<r_s$, $\partial/\partial r$ become time-like while
$\partial/\partial t$ become space-like. This means that inside
the black hole the metric has time-dependent coefficients. Thus,
one can use this correspondence to make a quantum theory of black
holes based on a quantized cosmological model.

In this letter we deal with the Hamiltonian formalism of a
time-dependent spherically symmetric space-time. We show that the
classical solutions of such a dynamical system can be identified
with the interior space-time of a Schwarzschild black hole. In
this model the metric functions play the role of independent
dynamical variables which with their conjugate momenta construct
the corresponding phase space. We then consider a Hamiltonian
quantum theory by replacing the classical phase space variables by
their Hermitian operators. We study the various aspects of the
resulting quantum model and corresponding  Wheeler-DeWitt (WD)
equation and present closed form expressions for the wavefunction
of the black hole. We shall see that the quantum solutions also
represent a quantization rule for the mass of the black hole. In
what follows, we work in units where $c=\hbar=1$.
\section{The model}
We start with the Einstein-Hilbert action
\begin{eqnarray}\label{B}
{\cal S}=\frac{1}{16\pi G}\int d^4x \sqrt{-g}{\cal R}+S_{YGH},
\end{eqnarray}
where ${\cal R}$ is the Ricci scalar, $g$ is the determinant of
the metric tensor and $S_{YGH}$ is the York-Gibbons- Hawking
boundary term . We assume that the geometry of space-time is
described by a spherically symmetric metric with time-dependent
line element as
\begin{eqnarray}\label{C}
ds^2=-\frac{N^2(t)}{\nu(t)}dt^2+\nu(t)dr^2+h^2(t)(d \vartheta^2+\sin^2
\vartheta d\varphi^2),
\end{eqnarray}
where $N(t)$ is the lapse function, while $\nu(t)$ and $h(t)$ are
functions of $t$ only, which play the role of our dynamical
variables to construct the phase space. It is clear that the above
metric with $t=\mbox{const.}$ and $r=\mbox{const.}$ describes the
surface of a $2$-sphere with radius $h(t)$ and area $A=4 \pi
h^2(t)$. By substituting (\ref{C}) into (\ref{B}) and integrating
over spatial dimensions, we are led to an effective Lagrangian in
the minisuperspace $(\nu,h)$ as
\begin{eqnarray}\label{E}
{\cal
L}=-\frac{V_0}{8\pi G}\left[\frac{1}{N}\left(h\dot{h}\dot{\nu}+\dot{h}^2\nu\right)-N\right],
\end{eqnarray}
in which $V_0$ is the volume of the space part of the action where
is treated to be a finite constant. A point $(\nu,h)$ in this
minisuperspace represents a $4$-geometry. Although, in this step
we can vary the above Lagrangian to get the equation of motion for
$h$ and $\nu$, but the Hamiltonian constraint resulting from this
Lagrangian does not have the desired form for construction of WD
equation describing the relevant quantum model. Thus, to transform
Lagrangian (\ref{E}) to a more manageable form, consider the
following change of variables
\begin{equation}\label{F}
x-y=h,
\hspace{0.5cm}
x+y=h\nu.
\end{equation}
In terms of these new variables, Lagrangian (\ref{E}) takes the form
\begin{eqnarray}\label{I}
{\cal
L}=-\frac{V_0}{8\pi G}\left[\frac{1}{N}\left(\dot{x}^2-\dot{y}^2\right)-N\right].
\end{eqnarray}
Now, if we use the following coordinate transformation which is
introduced in \cite{s1}
\begin{eqnarray}\label{mis2}
\begin{array}{cc}
(x-y)^{\frac{1}{2}}=\frac{1}{2}(u-v),\\
\\
(x+y)^{\frac{1}{2}}=\frac{1}{2}(u+v),
\end{array}
\end{eqnarray}
and the lapse rescaling
\begin{eqnarray}\label{mis1}
N=(x^2-y^2)^{\frac{1}{2}}\widetilde{N},
\end{eqnarray}
then the corresponding Lagrangian becomes
\begin{eqnarray}\label{Lag}
{\cal L}=-{M^2_{Pl}V_0}\widetilde{N}^{-1}(\dot{u}^2-\dot{v}^2)+{M_{Pl}^2V_0}\widetilde{N}(u^2-v^2),
\end{eqnarray}
which describes an isotropic oscillator-ghost-oscillator system.

To construct the Hamiltonian of the model, note that the momenta
conjugate to $u$ and $v$ are
\begin{eqnarray}\label{J}
\begin{array}{cc}
\Pi_u=-\frac{2V_0M^2_{Pl}}{\widetilde{N}}\dot{u},\\\\
\Pi_v=\frac{2V_0M^2_{Pl}}{\widetilde{N}}\dot{v}.
\end{array}
\end{eqnarray}
Also, the primary constraint is given by
\begin{eqnarray}\label{K}
\Pi_{\widetilde{N}}=\frac{\partial {\cal L}}{\partial
\dot{\widetilde{N}}}=0.
\end{eqnarray}
In terms of the conjugate momenta the Hamiltonian is given by
\begin{equation}\label{K1}
H=\dot{u}\Pi_u+\dot{v}\Pi_v-{\cal L},\end{equation}leading to
\begin{equation}\label{K2}
H=-\frac{\widetilde{N}}{4M^2_{Pl}V_0}(\Pi_u^2-\Pi_v^2)-V_0M^2_{Pl}\widetilde{N}(u^2-v^2).\end{equation}
Because of the existence of constraint (\ref{K}), the Lagrangian of the system is singular and
the total Hamiltonian can be constructed by adding to $H$ the primary constraints multiplied
by arbitrary functions of time $\lambda(t)$
\begin{eqnarray}\label{L}
H_T=-\frac{\widetilde{N}}{4M^2_{Pl}V_0}(\Pi_u^2-\Pi_v^2)-V_0M^2_{Pl}\widetilde{N}(u^2-v^2) +\lambda \Pi_{\widetilde{N}}.
\end{eqnarray}
The requirement that the primary constraint should
hold during the evolution of the system means that
\begin{eqnarray}\label{M}
\dot{\Pi}_N=\left\{\Pi_N,H_T\right\}\approx
0,\end{eqnarray}
which leads to the secondary (Hamiltonian)
constraint
\begin{eqnarray}\label{N}
{\cal H}=-\frac{1}{4V_0M_{Pl}^2}(\Pi_u^2-\Pi_v^2)-{V_0M_{Pl}^2}(u^2-v^2)\approx
0.
\end{eqnarray}
Before going any further, some remarks are in order about the mass
of the system. In the usual classical theory, the unique solution
to the vacuum Einstein equation for the spherically symmetric
space-time is the Schwarzschild solution in which there exists an
integration constant representing the mass of the black hole. In
the canonical formalism for the spherically symmetric space-time,
by following Fischler-Morgan-Polchinski \cite{ss1} and  Kuchar
\cite{ss2} calculations the mass of a black hole can be regarded
as a dynamical variable, which can be expressed as a function of
canonical data. The spherically symmetric hypersurface on which
the canonical data is given is supposed to be embedded in a
Schwarzschild black hole space-time whose metric is given by
\begin{eqnarray}\label{ss1}
N=1,\hspace{.5cm} \nu=\frac{2MG}{t}-1,\hspace{.5cm} h=t.
\end{eqnarray}
This identification of the space-time with the canonical data
enables us to connect the Schwarzschild mass $M$ with the
canonical data on any small piece of a space-like hypersurface. As
a result, from combination of the relations (\ref{F}),
(\ref{mis2}), (\ref{J}) and (\ref{ss1}) the mass function can be
expressed by the canonical variables as
\begin{eqnarray}\label{ss4}
M=\frac{\pi}{2V_0}\left(\frac{1}{4M^2_{Pl}V_0}\left(\Pi_u+\Pi_v\right)^2+V_0M^2_{Pl}(u-v)^2\right).
\end{eqnarray}
It is easy to show that the Poisson bracket of mass function with
the Hamiltonian vanishes strongly
\begin{eqnarray}\label{ss5}
\{M,{\cal H}\}=0,
\end{eqnarray}
which shows that the mass is a constant of motion.

\section{Classical solutions}
The setup for constructing the phase space and writing the
Lagrangian and Hamiltonian of the model is now complete. The
classical dynamics is governed by the variation of Lagrangian
(\ref{Lag}) with respect to $u$ and $v$, that is
\begin{equation}\label{2-1-1}
\ddot{u}+\widetilde{N}^2u=0,\hspace{0.5cm}
\ddot{v}+\widetilde{N}^2v=0.\end{equation}Up to this point
the model, in view of the concerning issue of time, has
been of course under-determined. Before trying to solve these
equations we must decide on a choice of time in the theory. The
under-determinacy problem at the classical level may be removed by
using the gauge freedom via fixing the gauge. A glance at the above
equations shows that a suitable gauge is $\widetilde{N}=M_{Pl}$ which results the following equations
\begin{equation}\label{2-1}
\ddot{u}+M^{2}_{Pl}u=0,\hspace{0.5cm}
\ddot{v}+M^{2}_{Pl}v=0.
\end{equation}
Choosing some parameters $\theta_1$ and $\theta_2$ as the integration constants, the solutions are obtained as
\begin{equation}\label{2-2}
u=A\cos(M_{Pl}t+\theta_1),\hspace{0.5cm}
v=B\cos(M_{Pl}t+\theta_2),
\end{equation}where $A$ and $B$ are constants to be determined later. Now, the above solutions
should satisfy the constraint of a vanishing Hamiltonian. Thus,
substitution (\ref{2-2}) into (\ref{N}) gives a relation between
the constants $A$ and $B$
\begin{equation}\label{2-2-1}
A=\pm B,\end{equation}
implying that we can rewrite the solutions (\ref{2-2}) as
\begin{equation}\label{2-2-2}
u=A\cos(M_{Pl}t+\theta_1),\hspace{0.5cm}
v=A\eta\cos(M_{Pl}t+\theta_2),\end{equation}where $\eta$ takes the
values $\pm 1$ according to the choices in (\ref{2-2-1}). Since in
the quantum version of the model we are interested in constructing
wave packets from the WD equation, we would like to obtain a
classical trajectory in configuration space $(u,v)$, where the
classical time $t$ is eliminated. This is because no such
parameter exists in the WD equation. It is easy to see that the
classical solutions (\ref{2-2-2}) may be displayed as the
following trajectories
\begin{eqnarray}\label{2-4}
u^2+v^2-2\eta\cos(\Delta\theta)uv-A^2\sin^2(\Delta\theta)=0,
\end{eqnarray}
where $\Delta\theta=\theta_1-\theta_2$.
Equation (\ref{2-4}) describes ellipses which their major axes make angle $\pi/4$ with the positive/negative
$u$ axis  according to the choices $\pm 1$ for $\eta$. Also, the eccentricity and the size of each trajectory are determined
by $\Delta\theta$ and $A$ respectively.

On the other hand, inserting solution (\ref{2-2}) into the mass function, we obtain
\begin{eqnarray}\label{ss6}
M=2\pi M^2_{Pl}A^2\sin(\frac{\Delta\theta}{2}).
\end{eqnarray}
To obtain the usual Schwarzschild solution, consider time
rescaling $Ndt=d\tau$, where from (\ref{mis1}) we have
$N=M_{Pl}(u^2-v^2)/4$.  Consequently, solutions (\ref{2-2}) give
\begin{eqnarray}\label{2-6}
\cos(2M_{Pl}t+\theta_1+\theta_2)=\frac{-8(\tau-\tau_0)}{A^2\sin\Delta\theta},
\end{eqnarray}
where $\tau_0$ is a constant of integration. From equations (\ref{F}) and (\ref{mis2}) one then has
\begin{eqnarray}\label{2-7}
\begin{array}{cc}
h=\begin{cases}-2\coth(\frac{\Delta\theta}{2})\tau, \hspace{.5cm}\eta=1,\hspace{.3cm}\Delta\theta<0, \\
2\tanh(\frac{\Delta\theta}{2})\tau, \hspace{.5cm}\eta=-1,\hspace{.3cm}\Delta\theta>0. \\
\end{cases}
\end{array}
\end{eqnarray}
and
\begin{eqnarray}\label{mis3}
\nu=\begin{cases}\cot^2(\frac{\Delta\theta}{2})\left(\frac{2GM}{\tau}\cot(\frac{\Delta\theta}{2})-1\right),\hspace{.5cm}
\eta=1,\hspace{.5cm}\Delta\theta<0, \\
\tan^2(\frac{\Delta\theta}{2})\left(\frac{2GM}{\tau}\tan(\frac{\Delta\theta}{2})-1\right),\hspace{.5cm}\eta=-1,
\hspace{.5cm}\Delta\theta>0, \\
\end{cases}
\end{eqnarray}
in which we put
\begin{eqnarray}\label{mis4}
\begin{array}{cc}

\tau_0=-\frac{\eta}{8}A^2\sin\Delta\theta.
\end{array}
\end{eqnarray}
Thus, in terms of the metric functions $(h,\nu)$ the classical trajectories (\ref{2-4}) take the form
\begin{eqnarray}\label{mis5}
\nu=\frac{1}{2}\left(\frac{A^2\sin^2\Delta\theta}{1-\eta\cos\Delta\theta}\right)\frac{1}{h}-
\left(\frac{1+\eta\cos\Delta\theta}{1-\eta\cos\Delta\theta}\right).
\end{eqnarray}

Therefore, the above Lagrangian formalism leads us to the following interior
Schwarzschild black hole metric if we assume $\Delta\theta=2n\pi-\pi/2$ when
$\eta=1$ and $\Delta\theta=2n\pi+\pi/2$ when $\eta=-1$, respectively
\begin{eqnarray}\label{V}
ds^2=-\left(\frac{2MG}{\tau}-1\right)^{-1}d\tau^2+\left(\frac{2MG}{\tau}-1\right)dr^2+\tau^2\left(d
\vartheta^2+\sin ^2 \vartheta d\varphi^2\right).
\end{eqnarray}
It is clear from the condition $\nu(\tau)>0$ that this solution is
only valid for $\tau<2MG$. The above metric has an apparent
singularity at $\tau=2MG$. This singularity is like the coordinate
singularity associated with horizon in the Schwarzschild or de
Sitter space-time, and as is well known, there are other
coordinate system for which this type of singularity is removed
\cite{1}. Another singularity associated with the metric (\ref{V})
is its essential singularity at $\tau=0$. As we know, in general
relativity, to investigate the types of singularities one has to
study the invariants characteristics of space time and to find
where these invariants become infinite so that the classical
description of space-time breaks down. In a 4- dimensional
Riemannian space-time there are $14$ independent invariants, but
to detecting the singularities it is sufficient to study only
three of them, the Ricci scalar ${\cal R}$, $R_{\mu \nu}R^{\mu
\nu}$ and the so-called Kretschmann scalar $R_{\mu \nu \sigma
\delta}R^{\mu \nu \sigma \delta}$. For the metric (\ref{V}) the
Kretschmann scalar reads
\begin{eqnarray}\label{W}
K=R_{\mu \nu \sigma \delta}R^{\mu \nu \sigma \delta}=\frac{48G^2
M^2}{\tau^6}.
\end{eqnarray}
Now, it is clear that the space-time describing by the metric
(\ref{V}) has an essential singularity at $\tau=0$, where can not
be removed by a coordinate transformation. Note that the
Schwarzschild manifold contains an anisotropic expanding universe,
the ''white hole'' portion of the extended geometry, and also an
anisotropic collapsing universe, the black hole interior as well.
In this paper we focus attention on the black hole interior
portion of the geometry, but all conclusions may be restated in
terms of the expanding white hole geometry due to the time
reversal symmetry of both Schwarzschild geometries.

\section{Dirac Observables}
As is well known general relativity is invariant under the group
of diffeomorphisms of the space-time manifold ${\cal M}$. The main
consequences of such a diffeomorphism invariance are that the
Hamiltonian can be expressed as a sum of constraints and any
observable must commute with these constraints. An observable is a
function on the constraint surface such that is invariant under
the gauge transformations generated by all of the first class
constraints. By a first class constraint we mean a phase space
function with the property that it has weakly vanishing Poisson
bracket with all constraints. As an example the momentum and
Hamiltonian constraints are always first class, see (\ref{K}) and
(\ref{N}). The Hamiltonian and momentum constraints in general
relativity are generators of the corresponding gauge
transformations, and so a function on the phase space is an
observable if has weakly vanishing Poisson brackets with the first
class constraints. To find gauge invariant observables, we can
proceed as follows. In Lagrangian (\ref{Lag}), the unconstrained
phase space $\Gamma$ is $\mathbb{R}^4$ with global canonical
coordinates $(x_i,\Pi_i)$, $i=1,2$, with Poisson brackets
$\{x_i,\Pi_j\}=\delta_{ij}$. Now, let us define on $\Gamma$ the
complex-valued functions
\begin{eqnarray}\label{4-1}
\begin{array}{cc}
C_i=M_{Pl}\sqrt{V_0}x_i+i\frac{1}{2M_{Pl}\sqrt{V_0}}\Pi_i,\\
\\
C^*_i= M_{Pl}\sqrt{V_0}x_i-i\frac{1}{2M_{Pl}\sqrt{V_0}}\Pi_i,
\end{array}
\end{eqnarray}
where $x_i=\{u,v\}$ and $\Pi_i=\{\Pi_u,\Pi_v\}$. The set
$S=\{C_i,C^*_i,1\}$ on $\Gamma$ is closed under the Poisson
bracket, $\{C_i,C^*_j\}=-i\delta_{ij}$ and every sufficiently
regular function on $\Gamma$ can be expressed in terms of the sums
and products of the elements of $S$. Hence, the Hamiltonian and
mass can be viewed as
\begin{eqnarray}\label{4-2}
{\cal H}=C^*_v C_v-C^*_u C_u,
\end{eqnarray}
and
\begin{eqnarray}\label{ss7}
M=\frac{\pi}{2V_0}\left(C_u-C^{*}_v\right)\left(C^{*}_u-C_v\right).
\end{eqnarray}
Therefore, the classical Poisson algebra generated by the elements of
$S$ is sufficiently large for describing the classical dynamics of the system.
In the next section we will use this algebra as the starting point for quantization of the model.
The classical dynamics of these variables is
\begin{eqnarray}\label{4-3}
\begin{array}{cc}
C_u=AV_{0}^{\frac{1}{2}}M_{Pl}^2e^{i(t+\theta_1)},\\
\\
C_v=\eta AV_{0}^{\frac{1}{2}}M_{Pl}^2e^{-i(t+\theta_2)}.
\end{array}
\end{eqnarray}
To find a set of constraints of motion, consider on $\Gamma$ the functions
\begin{eqnarray}\label{4-4}
\begin{array}{cc}
J_{0}=-\frac{1}{2}(C^*_v C_v+C^*_u C_u),\\
\\
J_{+}=C_v C_u,\\
\\
J_{-}=C^*_v C^*_u,
\end{array}
\end{eqnarray}
whose Poisson brackets form a closed algebra by
\begin{eqnarray}\label{4-5}
\begin{array}{cc}
\{J_{+},J_{-}\}=2i J_0,\\
\\
\{J_0,J_{\pm}\}=\mp i J_{\pm}.
\end{array}
\end{eqnarray}
The elements of this algebra have strongly vanishing Poison brackets with the
Hamiltonian.
Since the physical space is two dimensional, there will be at most two independent
constraints. On the constraint surface ${\cal H}=0$, the functions $J$'s
are not algebraically independent but satisfy the identity
\begin{eqnarray}\label{4-7}
J_0^2-J_+ J_-=0.
\end{eqnarray}
Also, we can write the mass function as a combination of $J$'s as
\begin{eqnarray}\label{ss8}
M=-\frac{\pi}{2V_0}(2J_0+J_{+}+J_{-}).
\end{eqnarray}
Thus, the quantity $M$ is a gauge invariant observable.

\section{Quantization of the model}
\begin{figure}
\centering
\includegraphics[width=8cm]{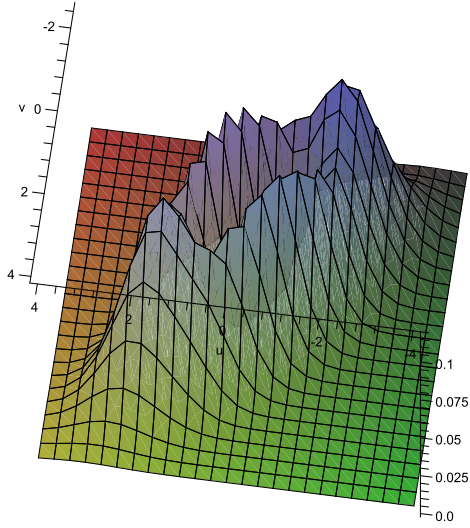}\\
\caption{The figure shows $|\Phi(x,y)|^2$ the square of the
wavefunction. This figure is plotted for numerical values
$\alpha=1$, $\delta\alpha=0.3$, $\beta=\pi/4$,
$\delta\beta=3\pi/20$, $\eta=1$ and $\omega=1$.}\label{fig1}
\end{figure}

We now focus attention on the study of the quantization of the
model described above. The quantum version of the model described
by relations $\{x_i,\Pi_j\}=\delta_{ij}$ can be achieved via the
canonical quantization procedure which leads the following
commutation relations
\begin{eqnarray}\label{5-1}
[u,\Pi_u]=i, \hspace{.5cm}[v,\Pi_v]=i.
\end{eqnarray}
Then, the set of hermitian quantum operators $\hat{S}=\{C_i,C_i^{\dagger},1\}$
will have the following commutator algebra
\begin{eqnarray}\label{5-2}
\begin{array}{cc}
[C_i,C_j^{\dagger}]=\delta_{ij}1,\hspace{.5cm}
[C_i,1]=[C^{\dagger}_i,1]=0.
\end{array}
\end{eqnarray}
The set $\hat{S}$ and its commutator algebra are the quantum counterpart
of the set $S$ and its Poisson bracket algebra.
The operator version of the classical Hamiltonian constraint takes the form
\begin{eqnarray}\label{5-3}
H=C_v^{\dagger}C_v-C_u^{\dagger}C_u.
\end{eqnarray}
Let us define vacuum state according to
\begin{eqnarray}\label{5-4}
C_u|0,0>=0,\hspace{.5cm}C_v|0,0>=0,
\end{eqnarray}
so that $C_i$ and $C_i^{\dagger}$ are annihilation and  creation operators respectively. Note
that by the Hamiltonian operator (\ref{5-3}), zero point energy is canceled out.
Consequently, the WD equation can be written as
\begin{eqnarray}\label{5-4a}
H\Psi(u,v)=\left(-\frac{\partial^2}{\partial u^2}+\frac{\partial^2}{\partial v^2}+\omega^2(u^2-v^2)\right)\Psi(u,v)=0,
\end{eqnarray}
where $\omega=V_0/8\pi G$. This equation is a quantum isotropic
oscillator-ghost-oscillator system with zero energy. Therefore,
its solutions belong to a subspace of the Hilbert space spanned by
separable eigenfunctions of a two-dimensional isotropic simple
harmonic oscillator Hamiltonian. Separating the eigenfunctions of
(\ref{5-4a}) in the form
\begin{eqnarray}\label{5-4b}
\Phi_{n_1,n_2}(u,v)=X_{n_1}(u)Y_{n_2}(v),
\end{eqnarray}
yields
\begin{eqnarray}\label{5-4c}
\begin{array}{cc}
X_{n_1}(u)=\left(\frac{\omega}{\pi}\right)^{1/4}\left[\frac{H_{n_1}(\omega^{1/2}u)}{\sqrt{2^{n_1}n_1!}}\right]e^{-\omega
u^2/2},\\
\\
Y_{n_2}(v)=\left(\frac{\omega}{\pi}\right)^{1/4}\left[\frac{H_{n_2}(\omega^{1/2}v)}{\sqrt{2^{n_1}n_2!}}\right]e^{-\omega
v^2/2}.
\end{array}
\end{eqnarray}
subject to the restriction $n_1= n_2 = n$. In (\ref{5-4c}), $H_n(x)$ are Hermite polynomials and
the eigenfunctions are normalized according to
\begin{eqnarray}\label{5-4d}
\int_{-\infty}^{+\infty}e^{-x^2}H_n(x)H_m(x)dx=2^n \pi^{1/2}n! \delta_{mn}.
\end{eqnarray}
The set $\{\Phi_{n_1,n_2}(u,v)\}$ forms a closed span of the zero
sector subspace of the Hilbert space $L_2$ of measurable
square-integrable functions on $\mathbb{R}^2$ with the usual inner
product defined as
\begin{eqnarray}\label{5-4e}
\int \Phi_{n_1,n_2}(u,v)\Phi_{m_1,m_2}(u,v)dudv=\delta_{n_1,m_1}\delta_{n_2,m_2},
\end{eqnarray}
that is, the orthonormality and completeness of the basis functions follow from those of the Hermite polynomials.
Therefore, in the position representation, we may write the general solution of the
WD equation as a superposition of the above eigenfunctions
\begin{eqnarray}\label{5-5}
\Psi(u,v)=\left(\frac{\omega}{\pi}\right)^{1/2}\sum_{n=0}^{\infty}\frac{c_n}{2^nn!}e^{-\frac{\omega}{2}(u^2+v^2)}
H_n(\sqrt{\omega}u)H_n(\sqrt{\omega}v).
\end{eqnarray}
where $c_n$ are a set of complex constants. In general, one of the most important features in quantum WD approach is the recovery of
classical solutions from the corresponding quantum model or, in other words, how can the
WD wavefunctions predict a classical model. In this approach, one usually constructs a
coherent wave packet with good asymptotic behavior in the minisuperspace, peaking in the
vicinity of the classical trajectory.
Therefore, for our subsequent analysis, by using the equality
\begin{eqnarray}\label{5-6}
\sum_{n=0}^{\infty}\frac{s^n}{n!}H_n(x)H_n(y)=\frac{1}{\sqrt{1-s^2}}\exp\left[\frac{2xys-s^2(x^2+y^2)}{2(1-s^2)}\right],
\end{eqnarray}
we can evaluate the sum over $n$ in (\ref{5-5}) and simple
analytical expression for this sum is found if we choose the
coefficients $c_n$ to be  $c_n=B2^n\tanh\zeta$, where $B$ and
$\zeta$ are arbitrary complex constants, which results in
\begin{eqnarray}\label{5-7}
\begin{array}{cc}
\Psi(x,y)={\cal N}\exp[-\frac{1}{4}\cos(2\beta)\cosh(2\alpha)(x^2+y^2-2\eta\tanh(2\alpha)xy)]\\
\\
\times\exp[-\frac{i}{4}\sinh(2\alpha)\sin(2\beta)(x^2+y^2-2\eta\coth(2\alpha)xy)].
\end{array}
\end{eqnarray}
In this expression $\alpha$ and $\beta$ are the real and imaginary
parts of $\zeta$ respectively, that is, $\zeta=\alpha+i\beta$,
$x=\sqrt{\omega}u$, $y=\sqrt{\omega}v$ and ${\cal N}$ is a
numerical factor. In figure \ref{fig1} we have plotted the square
of the wavefunction for typical values of the parameters in which
we have taken the following combination of the solutions
\begin{eqnarray}\label{5-8}
\Phi(x,y)=\Psi_{\eta,\alpha,\beta}(x,y)-\Psi_{\eta,\alpha+\delta\alpha,\beta+\delta\beta},
\end{eqnarray}
for some $\delta\alpha$ and $\delta\beta$ in the vicinity of
$\alpha$ and $\beta$. It is seen that a good correlation exists
between the quantum pattern shown in this figure and the classical
trajectories (\ref{2-4}) in configuration space $(u,v)$.

In continuation of this section we focus attention on the Dirac observables of the black hole and try to
build an algebra of physical operators, that is, a subalgebra of ${\cal
A}$ that would leave $V_{HH,0}$ invariant. To do this, notice
that the Poisson bracket algebra of the classical $J$'s can be promoted into
a commutator algebra version by setting
\begin{eqnarray}\label{5-9}
J_+=C_vC_u,\hspace{.5cm}J_-=C_v^{\dagger}C_u^{\dagger},\hspace{.5cm}J_0=-\frac{1}{2}\left(C_v^{\dagger}C_v+C_u^{\dagger}C_u+1\right),
\end{eqnarray}
so that the corresponding commutators are
\begin{eqnarray}\label{5-10}
[J_+,J_-]=2J_0,\hspace{.5cm}[J_0,J_{\pm}]=\mp J_{\pm}.
\end{eqnarray}
Equations (\ref{5-10}) are recognized as the commutators of the Lie algebra
of $SO(2,1)$. Since all these operators commute with Hamiltonian $H$,
we choose our physical operator algebra to be the algebra generated by the
set $\{J_{\pm},J_0,1\}$. Note that in analogy to the classical algebraic
identity (\ref{4-7}) the quantum $J$'s are not independent and satisfy the
following algebraic relation
\begin{eqnarray}\label{5-11}
J_0^2-\frac{1}{2}(J_+J_-+J_-J_+)+\frac{1}{4}=0.
\end{eqnarray}
Now, the action of the operators $\{C_i,C_i^{\dagger}\}$ on the states of the physical
Hilbert space is
\begin{eqnarray}\label{5-12}
\begin{array}{cc}
C_u|n_1,n_2>=\sqrt{n_1}|n_1-1,n_2>,\hspace{0.5cm}C_u^{\dagger}|n_1,n_2>=\sqrt{n_1+1}|n_1+1,n_2>,\\
\\
C_v|n_1,n_2>=\sqrt{n_2}|n_1,n_2-1>,\hspace{0.5cm}C_v^{\dagger}|n_1,n_2>=\sqrt{n_2+1}|n_1,n_2+1>,
\end{array}
\end{eqnarray}
so that with the help of (\ref{5-9}), one can put the action of $J$'s on the physical states as
\begin{eqnarray}\label{5-13}
\begin{array}{cc}
J_0|n_1,n_2>=-\frac{1}{2}(n_1+n_2+1)|n_1,n_2>,\\
\\
J_+|n_1,n_2>=\sqrt{n_1n_2}|n_1-1,n_2-1>,\\
\\
J_-|n_1,n_2>=\sqrt{(n_1+1)(n_2+1)}|n_1+1,n_2+1>,
\end{array}
\end{eqnarray}
with constraint $n_1=n_2$. Consequently the expectation value of
mass operator for the interior solution will be \footnote{Note
that according to the definition of volume $V_0=4\pi \int dr$,
it's dimension is length.}

\begin{eqnarray}\label{5-15}
<M>_n=\left(\frac{\pi}{V_0}\right)\left(n+\frac{1}{2}\right),
\end{eqnarray}which means that black hole is quantized in discrete states with
energies $E_n$, that is (in ordinary units)

\begin{equation}\label{5-16}
E_n=\frac{\pi c
\hbar}{V_0}\left(n+\frac{1}{2}\right).\end{equation} It is well
known that the parameter $M$ in the outside of a black hole is
measured at spatial infinity where the Newtonian approximation is
valid and so it is a measure of mass of the black hole. In this
sense, we know that the energy spectrum of a quantized black hole
from an outside observer point of view is (see \cite{Bek} and
\cite{15})
\begin{equation}\label{5-17}
E_n=M_{Pl}c^2\sqrt{n},\end{equation} which gives the following
relation for the difference between two nearby states
\begin{equation}\label{5-18}
\Delta E=\frac{M_{Pl}c^2}{\sqrt{n+1}+\sqrt{n}}.\end{equation}Now,
it is obvious that $\Delta E$ tends to zero as $n\rightarrow
\infty$ which is in agreement with the correspondence principle
and the black hole can be described classically in this regime. On
the other hand, the physical meaning of black hole mass parameter
may be different inside the event horizon. As equation
(\ref{5-16}) shows, the energy eigenvalues consist of equally
spaced spectrum which resemble the energy spectrum of the harmonic
oscillator. This is not surprising, since a decomposition of the
quantum fields inside the black hole into normal modes is
essentially a decomposition into harmonic oscillators that are
decoupled. Also, from (\ref{5-16}) we see even the lowest energy
level, i.e. the level $n=0$, has some nonzero energy, the so
called {\it ground state energy}. The existence of such a energy
level is a purely quantum mechanical effect and may be interpreted
in terms of the uncertainty principle. Indeed, it is this
zero-point energy that is responsible for the fact that the system
under consideration (interior of the black hole in our case) does
not "freeze" at extremely low energies. Another feature of the
result (\ref{5-16}) is its agreement with the correspondence
principle in the sense that $\frac{E_{n+1}}{E_n}\rightarrow 1$ as
$n\rightarrow \infty$. As the quantum number $n$ increases the
system tends to its classical regime which in this case is a
superposition of the eigenstates of the type (\ref{5-5}). Such a
so-called coherent state consists of unblurred wave packets which
minimize the uncertainty relation.

\section{Conclusions}
It is interesting to investigate how the interior of a black holes
would be quantized. Discrete spectra arise in quantum mechanics in
the presence of a periodicity in the classical system, which in
turn leads to the existence of an adiabatic invariant or action
variable. Boher-Somerfeld quantization implies that this adiabatic
invariant has an equally spaced spectrum in the semi-classical
limit. Using this approach one can determine the mass and the area
spectrum of the black holes in view of an outside observer
\cite{14}. In this paper we have evaluated analytically the mass
spectrum of the interior Schwarzschild black hole by implementing
WD equation. The corresponding phase space of the interior of a
Schwarzschild black hole is spanned by two dynamical variables and
their conjugate momenta. We have shown that the classical
Lagrangian of the model gives rise the interior of Schwarzschild
solution.  Then, we studied quantization of the model through WD
equation and by imposing suitable conditions on the solutions of
WD equation we have obtained the mass spectrum of the interior of
the black hole. Our result shows that the mass spectrum, not only
is discrete, but also is equally spaced.

\end{document}